# Influence of chemical kinetics on spontaneous waves and detonation initiation in highly reactive and low reactive mixtures

Mikhail Liberman [a] *, Cheng Wang [b], Chengeng Qian [b], JianNan Liu [b,c]

[a] *Nordic Institute for Theoretical Physics, Stockholm University, Roslagstullsbacken 23, 106 91 Stockholm, Sweden*

[b] *State Key Laboratory of Explosion Science and Technology, Beijing Institute of Technology, Beijing, 100081, China*

[c] *College of Mining Engineering, Taiyuan University of Technology, Taiyuan, 030024, China*

*Corresponding Author: mliber@nordita.org  (Mikhail Liberman)
Phone: +46855378444 (office)/ +46707692513 (mob)

# Influence of chemical kinetics on spontaneous waves and detonation initiation in highly reactive and low reactive mixtures

**Abstract**

Understanding the mechanisms of explosions is important for minimizing devastating hazards. Due to the complexity of real chemistry, a single-step reaction mechanism is usually used for theoretical and numerical studies. The purpose of this study is to look more deeply into the influence of chemistry on detonation initiated by a spontaneous wave. Results of high resolution simulations performed for one-step models are compared with simulations for detailed chemical models for highly reactive and low reactive mixtures. The calculated induction times for $H_2$/air and for $CH_4$/air are validated against experimental measurements for a wide range of temperatures and pressures. It is found that the requirements in terms of temperature and size of the hot spots, which produce a spontaneous wave capable to initiate detonation, are quantitatively and qualitatively different for one-step models compared to the detailed chemical models. The time and locations when the exothermic reaction affects the coupling between the pressure wave and spontaneous wave are considerably different for a one-step and detailed models. The temperature gradients capable to produce a detonation and the corresponding size of hot spots are much shallower and, correspondingly, larger than those predicted with one-step models. The impact of detailed chemical model is particularly pronounced for the methane-air mixture. In this case, not only the hot spot size is much greater than that predicted by a one-step model, but even at elevated pressure the initiation of detonation by a temperature gradient is possible only if the temperature outside the gradient is so high, that can ignite thermal explosion. The obtained results suggest that the one-step models do not reproduce correctly the transient and ignition processes, so that interpretation of the simulations performed using a one-step model for understanding mechanisms of flame acceleration, DDT and the origin of explosions must be considered with great caution.

(Word count: Abstract-301)

# 1. Introduction

Understanding the causes and mechanisms of explosions is essential for minimizing the explosion hazard in many industrial processes such as: coal mines and natural gas pipelines, hydrogen energy, nuclear, chemical and other industries [1-4]. Historically, explosions of methane/air mixtures are long time known to occur in coal mines [5] and the mining industry, have one of the highest injury and fatality rates. Understanding the causes and mechanisms of explosions is essential for improving safety measures and minimizing devastating hazards. In the worst case explosions may be accompanied by detonation resulting in a considerable pressure rise and serious damage. If ignited in a confined area (pipes, tunnel, etc.) the flame accelerates and may undergo the deflagration-to-detonation transition (DDT), which can present significant hazards. Since the discovery of the deflagration-to-detonation transition more than 150 years ago, a large number of experimental, theoretical and numerical studies have been undertaken in attempt to understand the fundamental mechanisms and processes which lead to DDT (see e.g. the reviews [6, 7]). However, despite many years of intensive research, there are still many questions that are poorly understood, and the mechanisms of flame acceleration and DDT remain one of the main challenges in the combustion physics.

Because of the complexity of real chemical kinetics, the majority of numerical simulations undertaken in an attempt to understand the nature of DDT used a simplified one-step Arrhenius model. To justify this approach, some authors (e.g. [8]) argued that: "for many practical situations, an extensive description of the details of the chemical pathways is unnecessary. Instead, it is more important to have an accurate model of the fluid dynamics coupled to a model for the chemical-energy release that puts the released energy in the ''right'' place in the flow at the ''right'' time". The conclusion derived from the simulations with a one-step model [6-8] was that that the accelerating flame causes the formation of hot spots in unreacted gas ahead of the flame, which can then produce a detonation through the Zel'dovich gradient mechanism involving gradients of reactivity [9, 10].

This trend was considered as the mainstream in DDT studies until it was shown experimentally by Kuznetsov et al. [11] that for a stoichiometric hydrogen/oxygen and ethylene-air mixtures the temperature in the vicinity of the flame prior to DDT remains too low (does not exceed 550K) for spontaneous ignition. Experimental studies and numerical simulations of DDT overtaken by Liberman et al. [12-15] using the detailed chemical model for hydrogen/oxygen have shown that the DDT mechanism is different from the gradient mechanism. A new mechanism of DDT consisting in the mutual amplification of a weak shock formed close ahead of the flame front and coupled with the flame reaction zone was proposed by Liberman et al. [12]. The scenario of the shock coupling and coherent amplification of the shock and the flame reaction resembles the SWACER mechanism (shock wave amplification by coherent energy release) considered by Lee and Moen [16]. However, the disadvantage of using a single-step model in numerical simulations leads not only to an incorrect interpretation of the DDT mechanism but also gives a wrong, about two times smaller, value of the run-up distance [12, 14, 15].

In the present paper we consider the influence of chemical kinetics on the modeling of detonation initiated by a temperature gradient in the highly reactive hydrogen/air and in the slow reactive methane/air mixtures. To do so, we compare and contrast the simulations for a one-step model with the simulations with the multi-step detailed chemical models. Since the use of detailed chemical mechanisms can severely limit the calculations, the reduced chemical models with a minimum number of reactions and species, suitable for describing transient combustion, such as DDT, are of great interest, especially for methane/air combustion where complete reaction mechanism can consist of many hundred species and thousands of reactions. Therefore, we use the reduced chemical models with a minimum number of reactions and species but sufficiently detailed to correctly describe ignition delay times and characteristics of laminar flames for a wide range of initial pressures and temperatures. It is shown that there is considerable difference in the steepness of temperature gradient and consequently in the hot spot size capable to initiate a detonation for a detailed chemical reaction model compared to a one-step chemical model. The size of a hot spot with a temperature gradient capable of producing detonation obtained with a detailed chemical model can be in orders of magnitude greater than that obtained from calculations using a one-step model, especially in the case of low reactive methane-air. The difference in the steepness of temperature gradients, and correspondingly in the hot spot size, capable of producing detonation for a one-step and for a detailed chemical model caused by two factors. First, the induction time for a one-step model calibrated in such a way that the model more or less correctly reproduces the speed and width of the laminar flame is 2-3 orders of magnitude smaller than the actual induction time calculated using a detailed chemistry and measured in experiments. Another difference is that for a one-step model reaction is exothermic from the very beginning, while chain branching reactions start with endothermic induction stage representing chain initiation and branching. Therefore, the gasdynamics is effectively "switched-off" during the induction stage. As a consequence, combustion regimes initiated by the temperature gradient require much shallower temperature gradients compared with those predicted by a one-step model. This means that the size of a hot spot with a temperature gradient capable of producing detonation obtained with a detailed chemical model is by orders of magnitude greater than that obtained from calculations using a one-step model.

**2. Chemical kinetics modeling**

We compare the one-step chemical models, which have been used for 2D simulations of the flame acceleration and DDT in hydrogen/air [6, 17] and in methane/air [8], with the detailed chemical mechanisms. It is known that to be able to predict correctly the ignition delay times, the reduced mechanisms must consist of at least ten reactions [18]. All the chemical models used in the present study were thoroughly tested and compared with the standard GRI Mech 3.0 [19] and validated against experimental measurements. The requirements to the chemical schemes imply their capability to reproduce correctly laminar flame speeds and structure and the ignition delay times (induction times). These requirements stem from the need of ensuring the controllable capability of characterizing ignition for transient combustion processes evolving over wide ranges of temperature during the flame acceleration and DDT. Finally, computational costs to use chemical schemes with DNS or LES solvers must be reasonable for the feasibility of multidimensional simulations.

*2.1. Hydrogen-air chemistry: a single-step and detailed models*

A single-step chemistry approach suggests that the complex set of reactions can be modeled by a one-step Arrhenius reaction. We will compare with the detailed chemical model the one-step Arrhenius model, which was used in [6, 17] for 2D simulations of the hydrogen/air flame acceleration and DDT in channel with obstacles at initial pressure $P = 1 \text{atm}$ and initial temperature, $T_0 = 293 \text{K}$:

$$W = A\rho Y \exp(-E_a / RT). \qquad (1)$$

Here $A = 6.85 \cdot 10^{12} \text{cm}^3 / (\text{g s})$ is the pre-exponential factor, $\rho_0 = 8.7345 \cdot 10^{-4} \text{g} / \text{cm}^3$ is the gas mixture density, $\gamma = C_P / C_V = 1.17$ is the ratio of specific heats, $E_a = 46.37 \, RT_0$ is the activation energy of the reaction, R is the universal gas constant.

The detailed mechanism chosen to model the hydrogen-air chemistry is the mechanism developed by Kéromnès et al. [20], which consists of 19 reactions and 9 species. An advantage of this mechanism is that it was extensively validated over a large number of experimental conditions, especially focused on high pressures [21], and it shows an excellent agreement between the modeling and experimental measurements of the flame velocity-pressure dependence over a wide range of pressures and equivalence ratios.

*2.2. Methane/air chemistry: a single-step and multi-step mechanisms*

We compare a one-step Arrhenius model used by Kessler et al. [8] for 2D simulations of the methane/air flame acceleration and DDT in channel with obstacles at initial pressure $P = 1$ atm and temperature, $T_0 = 298\,K$. The model [8] was calibrated to give a reasonable approximation of the key properties of the methane/air that are: the laminar flame speed, the adiabatic flame temperature, the viscosity, and the speed of sound. The equation of a one-step kinetics is the same as Eq. (1) with the parameters for methane/air: $E_a = 67.55\,RT_0$, $A = 1.64 \cdot 10^{13}\,cm^3/(g\,s)$, $\rho = 1.1 \cdot 10^{-3}\,g/cm^3$, $\gamma = C_P/C_V = 1.197$.

The multi-step detailed mechanism chosen to model methane-air chemistry is the reduced detailed reaction model DRM-19 developed by Kazakov and Frenklach [22], which consists of 19 species and 84 reactions. The mechanism DRM-19 has been chosen for simulating as it was extensively validated by many researchers for combustion characteristics of $CH_4$/air related to ignition delay times and laminar flame velocities over a wide range of pressures, temperatures, and equivalence ratios [23, 24, 25].

**3. Comparison of one-step and detailed chemical schemes; induction times**

We compare predictions of one-step models with that obtained using detailed multi-step mechanisms. The ability of different reaction schemes to reproduce the laminar flame structure and speeds and the ignition delay times were examined using high resolution simulations. The resolution and convergence (a grid independence) tests were thoroughly performed to ensure that the resolution is adequate to capture details of the problem in question and to avoid computational artifacts. The convergence of the solution is quite satisfactory already for 8 grid points per flame width at initial pressure 1atm, but effective resolution up to 64 grid points and more was used for elevated pressures with corresponding cell sizes less 1μm (see Appendix).

The ignition delay times were calculated for different chemical reaction schemes using the standard constant volume adiabatic model. The ignition delay time can be defined as the time during which the maximum rate of temperature rise ($\text{Max}\{dT/dt\}$) is achieved, which is close to the time of exothermic reactions activation. While one-step models allow to reproduce the laminar flame speed $U_L$ and the adiabatic flame temperature $T_b$ with satisfactory good accuracy, the induction times predicted by the one-step models are significantly shorter than the induction times calculated using detailed chemical models, GRI 3.0 Mech, and the experimentally measured induction times.

*3.1. Hydrogen-air: one-step and detailed chemical mechanisms*

Figures 1(a, b) show the induction times versus temperature for hydrogen-air at pressures $P = 1\text{atm}$ and 2atm computed using the one-step model Eq.(1), the induction times computed with the detailed chemical model [20] and experimental results [25, 26, 29, 30, 31].

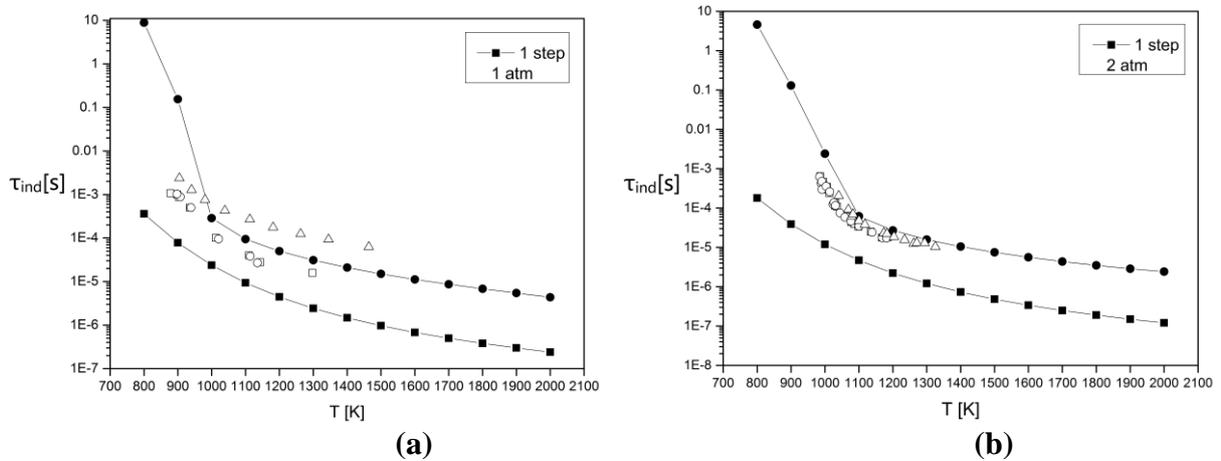

(a) (b)

**Figure 1**: Induction times for stoichiometric hydrogen-air mixture at pressure: $P = 1\text{atm}$ (a) and $P = 2\text{atm}$ (b). one-step model (■), detailed model [20] (•). Experiments (a): □- Snyder et al. [26]; ○- Slack and Grillo [27]; △ − Hu et al. [28]; (b): □ - Slack and Grillo [27]; ○- Slack [25]; △- Bhaskaran and Gupta. [29].

The induction times for hydrogen-air at pressures $P = 5\text{atm}$ and 10 atm computed using the one-step model Eq. (1) and using the detailed chemical model [20] and experimental results from Ref. [24, 28] are shown in Figs. 2 and 3. The empty squares and circles in Figs.2 and 3 are experimental results from Ref. [24, 28]. It is seen that the induction times predicted by the detailed chemical model [20] are in a good agreement with the experimental results but differ up to three orders of magnitude from that predicted by the one-step model. Another feature of the "real" induction time is an abrupt change of $(d\tau_{ind}/dT)$ at the crossover temperatures, which correspond to the transition from the endothermal induction stage to the exothermal stage. The crossover temperatures in Figs.1-3 are: $T_{cr} = 1000, 1100, 1200, 1300 K$ for $P = 1, 2, 5, 10 atm$, correspondingly.

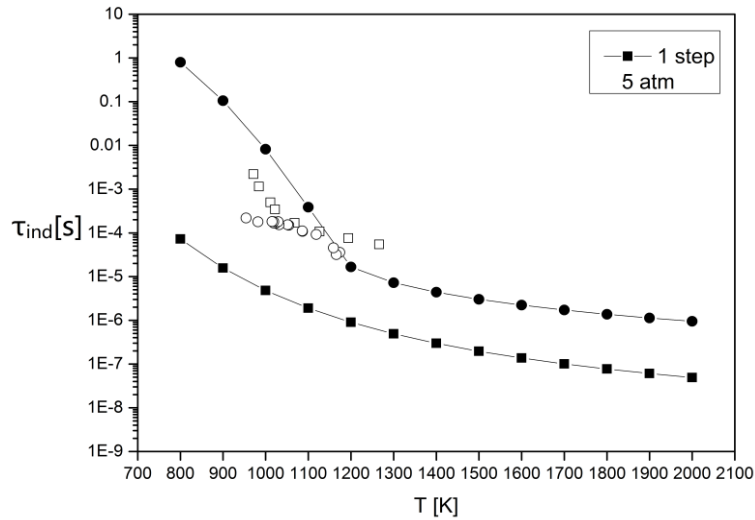

**Figure 2**: Induction times for stoichiometric hydrogen-air mixture at $P = 5 atm$ calculated for the one-step and detailed chemical models. Experimental data are: □- Hu et al. [28]; ○- Wang et al. [30]. In Fig. 2(b) experimental data are: □- Hu et al. [30]; ○- Pan et al. [31].

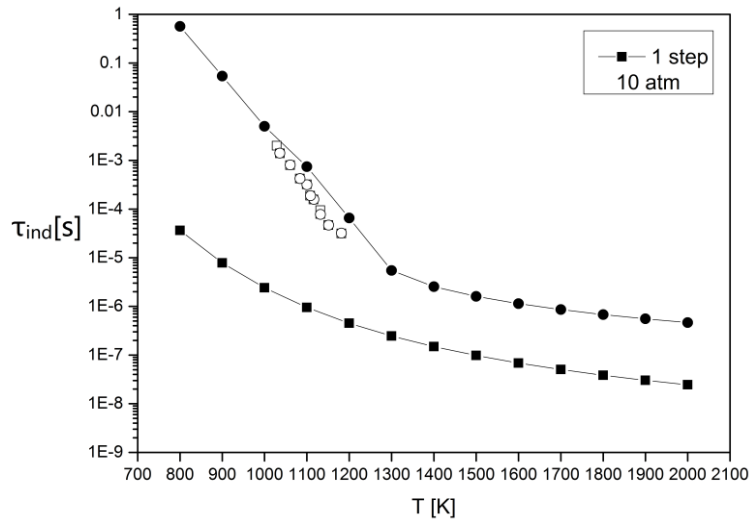

**Figure 3**: Induction times for stoichiometric hydrogen-air mixture at pressures $P = 10 atm$ calculated for the one-step and detailed chemical models. Experimental data are: □- Hu et al. [28], ○- Pan et al. [31].

### 3.2. Methane-air: a one-step and detailed chemical models

Figure 4 shows the induction times versus temperature at $P = 1 atm$, computed using the one-step model, the detailed chemical model DRM-19, GRI 3.0 Mech and the experimental measurements [32, 33].

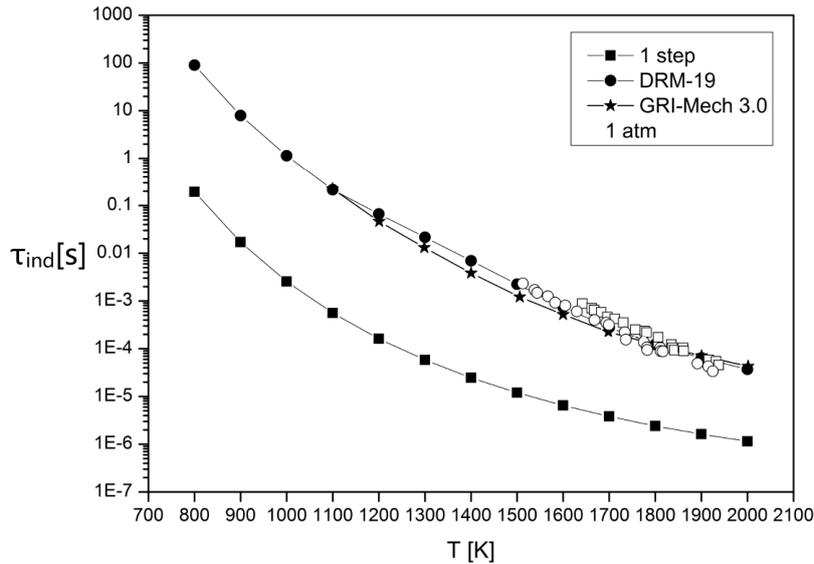

**Figure 4:** Induction times for methane-air at $P = 1\,\text{atm}$ calculated for the one-step model, DRM19 and GRI3.0. Empty circles and squares are experimental data Zeng et al. [32] and Hu et al. [33], correspondingly.

One can see that the difference between the induction time calculated with the one-step model and the real induction time calculated with detailed chemical models for methane/air is about 10 times larger than they are for hydrogen/air. A similar large difference between the induction times predicted by a one-step model and the induction times computed using detailed chemical models remains at elevated pressures as it is shown in Figs. 4 and 5 for the initial pressure 5 atm and 10 atm.

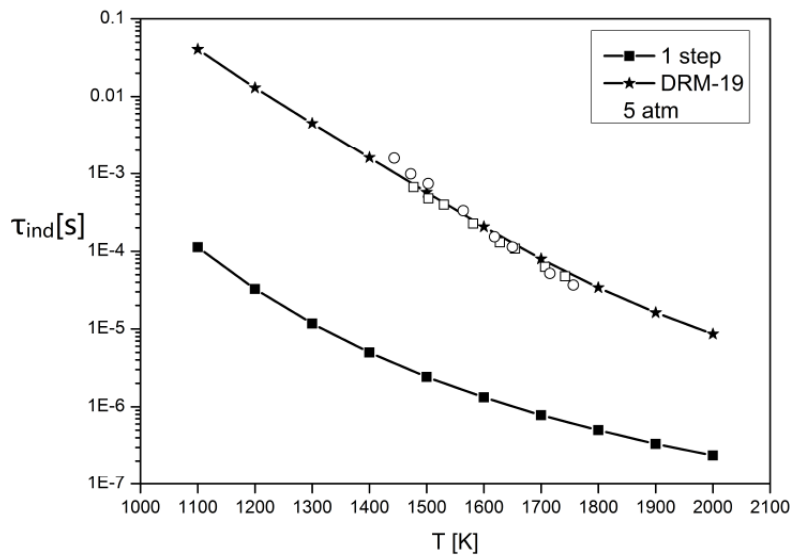

**Figure 5:** Induction times for methane-air at $P = 5\,\text{atm}$. Empty circles and squares are experimental results from [32] and [33], correspondingly.

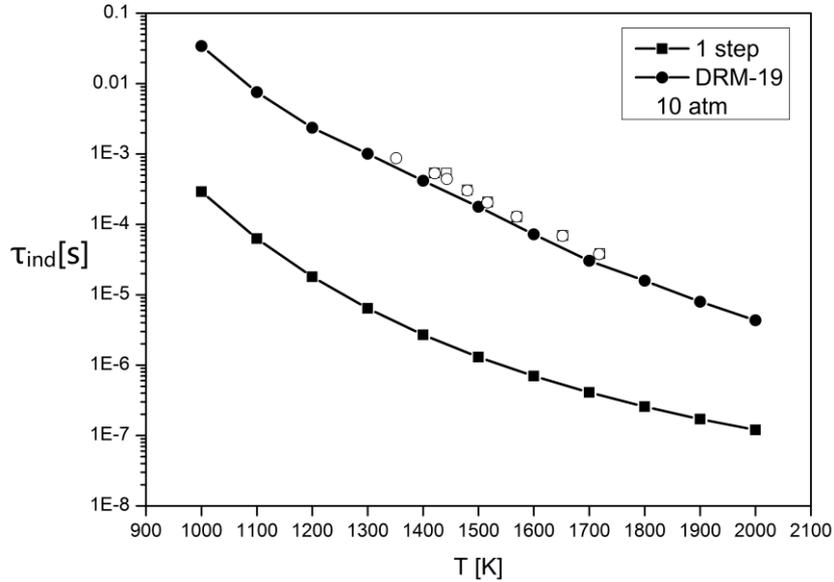

**Figure 6**: Induction times for methane-air at $P = 10\,\text{atm}$. Empty circles and squares are experimental results from [32] and [33], correspondingly.

It should be noted that, contrary to hydrogen/air, the temperature dependence of "real" induction time for methane/air does not show the transition corresponding to the crossover temperature - an abrupt change in $(d\tau_{ind}/dT)$ as it is for hydrogen/air.

**4. Spontaneous waves and detonation initiation by temperature gradients**

We will use the conventional term a "hot spot", which is an area within a reactive mixture, where temperature is higher than in surrounding mixture. The hot spot size is considered as the temperature gradient scale, $L = T/(dT/dx)$. Typically turbulence time scales are much longer than the induction times for temperatures higher than the ignition threshold temperature, which is defined as the temperature at which the fresh mixture ignites before it is consumed by the flame burned by the flame during $\tau_f = L_f/U_f$.

*4.1. Spontaneous waves and detonation initiation by a temperature gradient*

The ignition of a flammable mixtures is one of the most important and fundamental problems in combustion physics. In practical cases ignition begins in a small area of combustible mixture, which is locally heated by means of an electric spark, hot wire, and the like, and combustion begins in the form of a deflagration mode. Such local energy release results in the formation of an initially nonuniform distribution of temperature (or concentration of reagents), which depending on the mixture reactivity and the initial pressure determines the evolution of the reaction wave. One needs to know how the initial conditions in such "hot spot" influence the regime of the reaction wave, which is ignited and propagates out from the ignition location. The question of how a hot spot can give rise to different combustion modes remained open until Zel'dovich et al. [9] have demonstrated that a sufficiently shallow temperature gradient in the hot spot can initiate a detonation. The Zel'dovich's concept [10] of the spontaneous reaction wave propagating through a reactive mixture along a spatial gradient of reactivity is of great fundamental and practical importance. It opens an avenue to study ignition of different regimes of the reaction wave initiated by the initial non-uniformity in temperature or reactivity caused by the local energy release [34].

In a region with nonuniform distribution of temperature the reaction begins at the point of minimum ignition delay time and, correspondingly, the maximum temperature, and then it spreads along the temperature gradient by spontaneous autoignition at neighboring locations where $\tau_{ind}$ is longer. In the case of a one-step chemical model the induction time is defined by the time-scale of the maximum reaction rate. For a detailed, chain branching chemistry this is the time scale of the stage when the endothermic chain initiation stage completed and branching reactions begin. In the case of a one dimensional problem the spontaneous autoignition wave propagates relative to the unburned mixture in the direction of temperature gradient with the velocity:

$$U_{sp} = \left|(d\tau_{ind}/dx)\right|^{-1} = \left|(\partial\tau_{ind}/\partial T)^{-1}(\partial T/\partial x)^{-1}\right| \qquad (2)$$

Since there is no causal link between successive autoignitions, there is no restriction on the value of $U_{sp}$, which depends only on the steepness of temperature gradient and $\partial\tau_{ind}/\partial T$. It is obvious, that a very steep gradient (hot wall) ignites a deflagration mode (flame), while a zero gradient corresponds to uniform thermal explosion, which occurs in the induction time. The velocity of spontaneous wave initiated by the temperature gradient decreases while the autoignition wave propagates along the gradient, and reaches the minimum value at the point close to the cross-over temperature, where it can be caught-up and coupled with the pressure wave, which was generated due to the chemical energy release behind the high-speed spontaneous wave front. As a result, the pressure peak is formed at the reaction front, which grows at the expense of energy released in the reaction. After the intersection of the spontaneous wave front and the pressure wave, the spontaneous wave transforms into combustion wave and the pressure wave steepens into the shock wave. After the pressure peak becomes large enough, it steepens into a shock wave, forming an overdriven detonation wave. Classification of combustion regimes initiated by a temperature gradient for a one-step chemical model have been made by Zel'dovich [10], see also the review paper by Kapila et al [35].

Liberman et al. [36, 37] employed detailed chemical kinetic model to study the combustion regimes initiated by the temperature gradient in stoichiometric hydrogen-oxygen and hydrogen-air mixtures. It was shown that the temporal evolution of a spontaneous wave calculated for the detailed kinetic model differs considerably form the predictions obtained from simulations with a one-step model. The difference in the steepness of temperature gradients, and correspondingly in the hot spot size capable of producing detonation for the one-step and the detailed chemical models caused by two main reasons. First, the induction time for a one-step model calibrated in such a way that the model more or less correctly reproduces the speed and width of the laminar flame is 2-3 orders of magnitude smaller than the actual induction time calculated using a detailed chemistry and measured in experiments. Another difference is that for a one-step model reaction is exothermic for all temperatures, while chain branching reactions start with endothermic induction stage representing chain initiation and branching. Therefore, the gasdynamics is effectively "switched-off" during the induction stage. As a consequence, combustion regimes initiated by the temperature gradient require much shallower temperature gradients compared with those predicted by a one-step model. This means that the size of a hot spot with a temperature gradient capable of producing detonation obtained with a detailed chemical model is much larger than that obtained from calculations using a one-step model. The size of a hot spot with a temperature gradient capable of producing detonation decreases with the increase of initial pressure, and may become of the order of few millimeters at very high pressures [37, 38].

*4.2. Problem setup*

We consider uniform initial conditions apart from a linear temperature gradient. The model of the linear temperature gradient is convenient for analysis and it has been widely used in many previous studies [9, 10, 35-43]. The initial conditions at $t=0$, prior to ignition are constant pressure and zero velocity of the unburned mixture. At the left boundary at $x=0$ the conditions are for a solid reflecting wall, where $u(0,t)=0$ and the initial temperature, $T=T*$ exceeds the ignition threshold value. Thus, the initial conditions are quiescent and uniform, except for a linear gradient in temperature (and hence density):

$$T(x,0) = T*-(T*-T_0)(x/L), \ 0 \leq x \leq L \qquad (3)$$

$$P(x,0) = P_0, \ u(x,0) = 0. \qquad (4)$$

The initial temperature gradient is characterized by the temperature $T(0,0)=T*$ at the left end, by the mixture temperature outside the gradient, $T(x \geq L, 0) = T_0$ and by the gradient steepness, $(T*-T_0)/L$. The "length" $L$, which characterizes the gradient steepness can be viewed as the hot spot size, where the initial temperature gradient was formed.

The 1D direct numerical simulations are performed to solve the set of the one-dimensional time-dependent, fully compressible reactive Navier-Stokes equations and chemical kinetics.

$$\frac{\partial \rho}{\partial t} + \frac{\partial (\rho u)}{\partial x} = 0, \qquad (5)$$

$$\frac{\partial Y_i}{\partial t} + u\frac{\partial Y_i}{\partial x} = \frac{1}{\rho}\frac{\partial}{\partial x}\left(\rho D_i \frac{\partial Y_i}{\partial x}\right) + \left(\frac{\partial Y_i}{\partial t}\right)_{ch}, \tag{6}$$

$$\rho\left(\frac{\partial u}{\partial t} + u\frac{\partial u}{\partial x}\right) = -\frac{\partial P}{\partial x} + \frac{\partial \sigma_{xx}}{\partial x}, \tag{7}$$

$$\rho\left(\frac{\partial E}{\partial t} + u\frac{\partial E}{\partial x}\right) = -\frac{\partial (Pu)}{\partial x} + \frac{\partial}{\partial x}(\sigma_{xx} u) + \frac{\partial}{\partial x}\left(\kappa(T)\frac{\partial T}{\partial x}\right) +$$

$$+ \sum_k \frac{h_k}{m_k}\left(\frac{\partial}{\partial x}\left(\rho D_k(T)\frac{\partial Y_k}{\partial x}\right)\right), \tag{8}$$

$$P = R_B T n = \left(\sum_i \frac{R_B}{m_i} Y_i\right)\rho T = \rho T \sum_i R_i Y_i, \tag{9}$$

$$\varepsilon = c_v T + \sum_k \frac{h_k \rho_k}{\rho} = c_v T + \sum_k h_k Y_k, \tag{10}$$

$$\sigma_{xx} = \frac{4}{3}\mu\left(\frac{\partial u}{\partial x}\right). \tag{11}$$

We use conventional notations: $P$, $\rho$, $u$, are pressure, mass density, and flow velocity, $Y_i = \rho_i/\rho$ - the mass fractions of the species, $E = \varepsilon + u^2/2$ - the total energy density, $\varepsilon$ - the inner energy density, $R_B$ - is the universal gas constant, $m_i$ - the molar mass of i-species, $R_i = R_B/m_i$, $n$ - the molar density, $\sigma_{ij}$ - the viscous stress tensor, $c_v = \sum_i c_{vi} Y_i$ - is the constant volume specific heat, $c_{vi}$ - the constant volume specific heat of i-species, $h_i$ - the enthalpy of formation of i-species, $\kappa(T)$ and $\mu(T)$ are the coefficients of thermal conductivity and viscosity, $D_i(T)$ - is the diffusion coefficients of i-species, $(\partial Y_i/\partial t)_{ch}$ - is the variation of i-species concentration (mass fraction) in chemical reactions.

The equations of state for the reactive mixture and for the combustion products were taken with the temperature dependence of the specific heats and enthalpies of each species borrowed from the JANAF tables (Joint Army Navy NASA Air Force Thermochemical Tables) and interpolated by the fifth-order polynomials [44]. The ideal gas equation of state was used in the case of a single-step model [6-8, 17]. The viscosity and thermal conductivity coefficients of the mixture were calculated from the gas kinetic theory using the Lennard-Jones potential [45]. Coefficients of the heat conduction of i-th species $\kappa_i = \mu_i c_{pi} / \mathrm{Pr}$ are expressed via the viscosity $\mu_i$ and the Prandtl number, $\mathrm{Pr} = 0.75$.

## 5. Detonation initiation by a temperature gradient in stoichiometric $H_2$/air mixture

The time evolution of spontaneous wave and the detonation initiation by the steepest temperature gradient in $H_2$/air mixture and therefore by the minimum size of the hot spot at the initial conditions $P_0 = 1\,\mathrm{atm}$, $T^* = 1500\,\mathrm{K}$, $T_0 = 300\,\mathrm{K}$, computed for a one-step model and with the detailed chemical reaction model [20] are shown in Figs. 7 and 8, correspondingly.

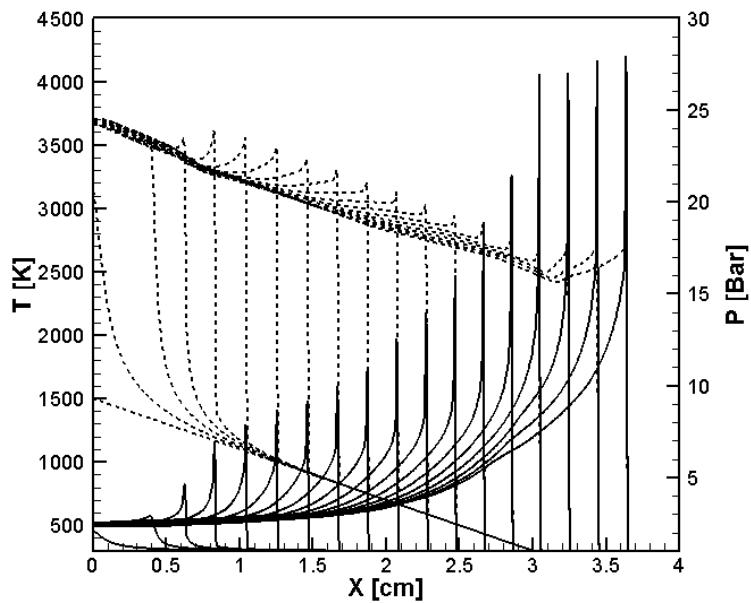

**Figure 7:** Time evolution of the temperature (dashed lines) and pressure (solid lines) profiles during detonation initiation in $H_2$/air for one-step model. $P_0 = 1\,\mathrm{atm}$, $\Delta t = 2\,\mu\mathrm{s}$.

The spontaneous reaction wave starts at the upper point of the gradient, where the temperature is maximum, and further its velocity decreases. If the gradient is sufficiently shallower, such that the minimum speed of the spontaneous wave is close to the sound speed, $a_s(T^*_{cr})$, the spontaneous reaction wave is coupling with the pressure pulse produced by the energy released in the reaction. As a result, a pressure peak is formed at the reaction front, which grows at the expense of energy released in the reaction. After the pressure peak becomes large enough, it steepens into a shock wave, forming an overdriven detonation wave. For a steeper temperature gradient (smaller hot spots) the velocity of the spontaneous wave at the minimum point is not sufficient to sustain synchronous feedback amplification between the reaction and the pressure pulse. In this case the pressure waves run ahead of the reaction wave out of the gradient and the result will be a deflagration or fast deflagration behind the shocks wave.

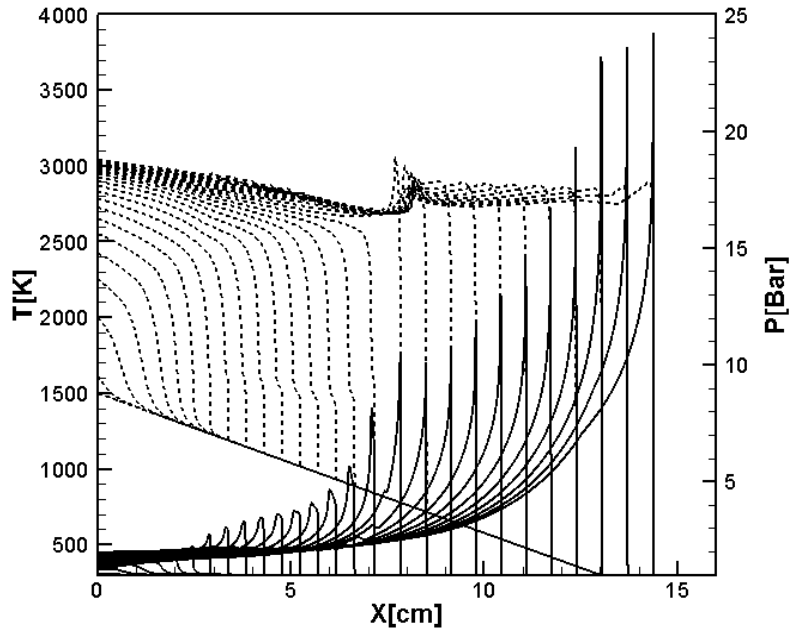

**Figure 8:** Time evolution of the temperature (dashed lines) and pressure (solid lines) profiles during detonation initiation in $H_2$/air calculated for the detailed chemical model [20]; $P_0 = 1\,\text{atm}$, $T^* = 1500\,\text{K}$, $\Delta t = 2\,\mu s$.

The velocity of the spontaneous wave initiated by the initial temperature gradient decreases along the gradient (since $\partial \tau_{ind} / \partial T$ decreases with increasing temperature, see Eq. (2)) and reaches its minimum value at the point close to the crossover temperature. Therefore, the necessary condition for initiating detonation by the spontaneous reaction wave is that the spontaneous wave initiated by the initial temperature gradient can be caught up and coupled with the pressure wave, which was generated behind the high-speed spontaneous wave front. Since the exothermic stage of the reaction begins and produces pressure pulse at the temperature slightly higher than the crossover temperature, the necessary condition for triggering a detonation can be written in the form

$$U_{sp}(T_{cr}^*) = \left(\frac{\partial \tau}{\partial T}(T_{cr}^*)\right)^{-1} \left(\frac{\partial T}{\partial x}(T_{cr}^*)\right)^{-1} = \left(\frac{\partial \tau}{\partial T}(T_{cr}^*)\right)^{-1} \frac{L}{T^* - T_0} \geq a_s(T_{cr}^*), \quad (12)$$

where the temperature $T_{cr}^*$ is slightly above the crossover temperature $T_{cr}$, and $a_s(T_{cr}^*)$ is the sound speed at the point corresponding to $\min\{U_{sp}\}$. Using this condition we can estimate the minimum size of the hot spot for the successful detonation initiation. Figures 9(a) and 9(b) show velocities of spontaneous wave at the minimum point as the function of the hot spot size calculated for the detailed and one-step chemical models at initial pressure $P_0 = 1$, 5 and 10atm; $T^* = 1500$ K, $T_0 = 300$ K.

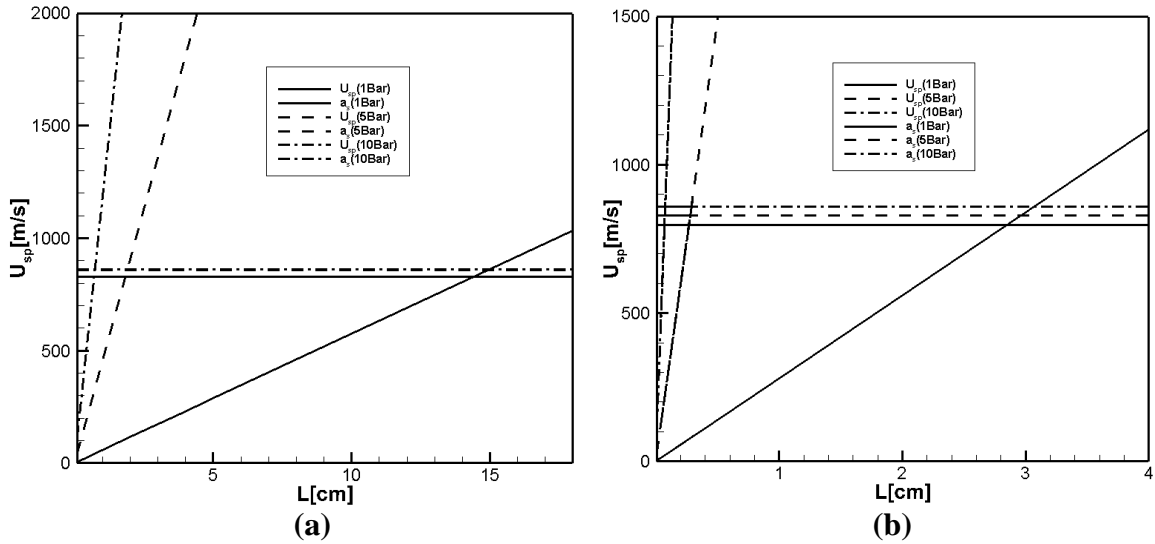

**Figure 9:** The minimum hot spot size $L = L_{cr}$ producing detonation at $P_0 = 1$, 5, 10atm. (a): detailed model, $T_{cr}^*(1\text{atm}) = 1300$ K, $T_{cr}^*(5\text{atm}) = 1400$ K, $T_{cr}^*(10\text{atm}) = 1410$ K. (b): one-step model $T_{cr}^*(1\text{atm}) = 1200$ K, $T_{cr}^*(5\text{atm}) = 1300$ K, $T_{cr}^*(10\text{atm}) = 1400$ K.

While at the normal or lower pressures the induction stage is much longer than the chain termination exothermic stage, they become of the same order at high pressures, when triple collisions dominate. The crossover temperature corresponding to the equilibrium of the induction and termination stages known as the extended second explosion limit [37] shifts to higher temperatures (Figs. 1-3) at high pressures. Since at high pressures the induction time decreases (more precisely, $\partial \tau_{ind} / \partial T$ decreases), the minimal steepness of the gradients necessary for detonation initiation increases. The corresponding size of the hot spot $L = L_{cr}$ producing detonation decreases. Figures 10(a, b) and 11(a, b) show the time evolution of temperature and pressure profiles during the initiation of detonation computed for the detailed [20] and one-step model at initial pressures $P_0 = 5\,\text{atm}$ and $P_0 = 10\,\text{atm}$.

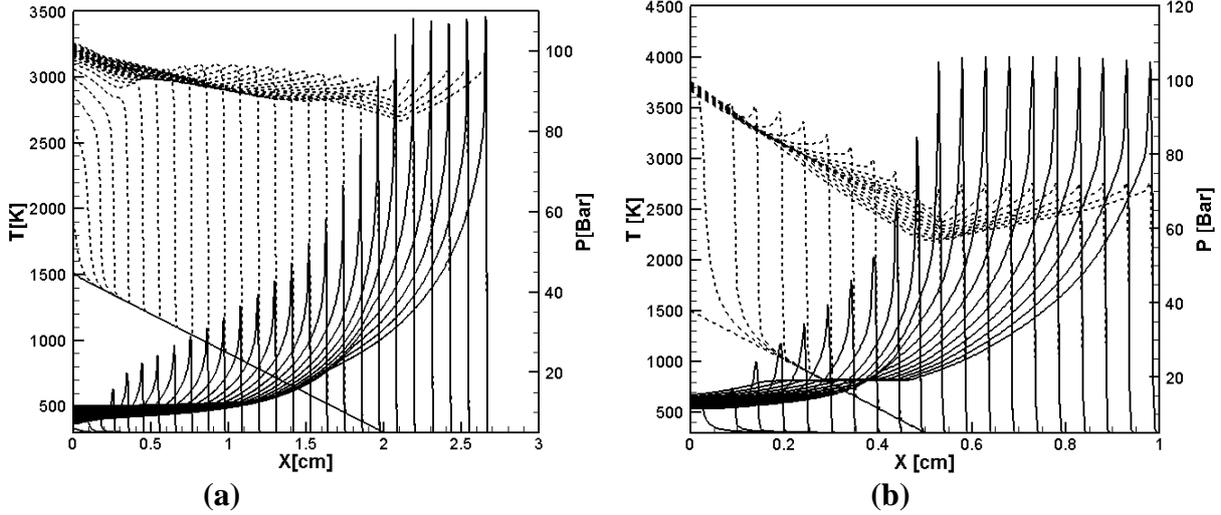

**Figure 10:** Time evolution of the temperature (dashed lines) and pressure (solid lines) profiles during detonation initiation in H$_2$/air at $P_0 = 5\,\text{atm}$, $T^* = 1500\,\text{K}$. (a): detailed model; (b): one-step model, $\Delta t = 2\,\mu s$.

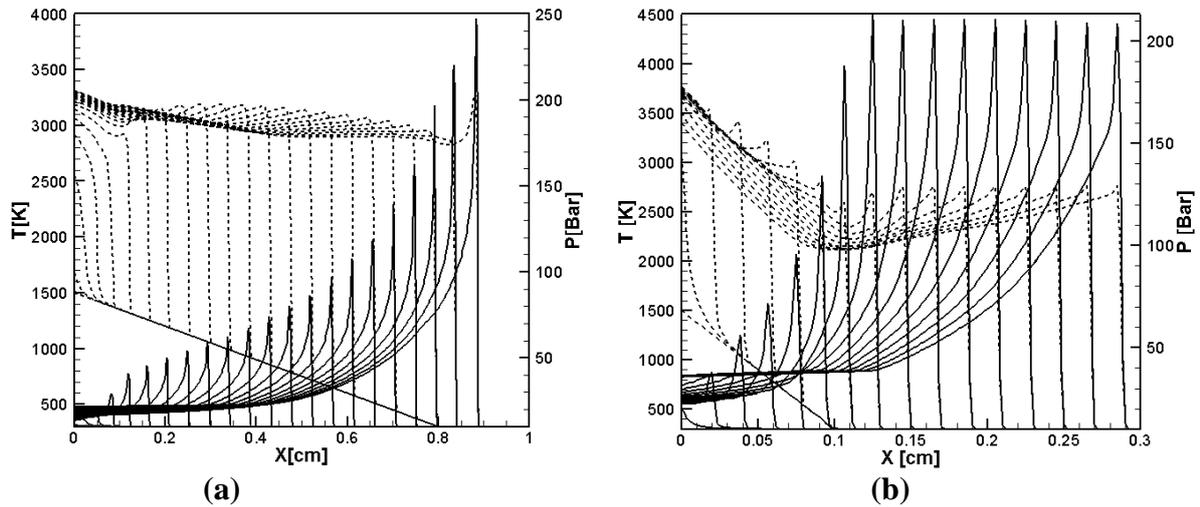

**Figure 11:** Time evolution of the temperature and pressure profiles during detonation initiation in H$_2$/air at $P_0 = 10\,\text{atm}$, $T^* = 1500\,\text{K}$. (a): detailed model; (b): one-step model.

By comparing the critical size of the hot spot, obtained in numerical simulations in Figs. 7, 8, 10(a, b) and 11(a, b), it can be seen that the equation (12) and diagrams in Figs. 9(a, b) predict the critical size with a good accuracy for both the detailed and one-step models over a wide range of pressures. Since during the induction stage there are no gasdynamic perturbations, the reaction proceed without heat release and the wave of exothermal reaction follows the spontaneous wave path with the delay determined by the duration of initiation-termination reactions. Therefore, the critical size of the hot spot predicted for the detailed chemistry is much larger than that for the one-step model for all pressures.

The difference in the sizes of hot spots, $L_{cr}$, at which the temperature gradient can produce a steady detonation for one-step and detailed models is due to the difference of $(\partial \tau_{ind} / \partial T)$ for these models, which determines the speed of the spontaneous wave (see Eq.(2)). According to Eq. (12) the ratio of critical sizes given by the detailed and one-step models for the same initial conditions can be estimated as

$$L_{det} / L_{1-step} \approx (\partial \tau_{ind} / \partial T)_{det} / (\partial \tau_{ind} / \partial T)_{1-step}. \tag{13}$$

Figure 12 shows $\partial \tau_{ind} / \partial T$ calculated for the detailed model [20] and for the one-step model [17]. It is seen that the difference between $(\partial \tau_{ind} / \partial T)_{det}$ and $(\partial \tau_{ind} / \partial T)_{1-step}$ remains approximately unchanged for all pressures.

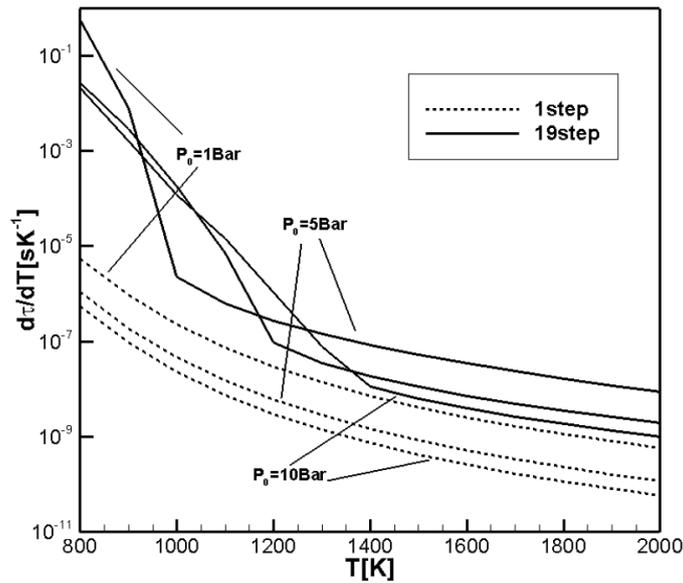

**Figure 12**: $(\partial \tau_{ind} / \partial T)$ for $H_2$/air at initial pressure 1, 5 and 10 atm calculated for detailed (solid lines) and one-step (dashed lines) models.

## 6. Spontaneous wave and detonation initiation by temperature gradient in CH$_4$/air

The induction times and $(\partial \tau_{ind}/\partial T)$ are much longer for methane/air at all temperatures compared to the hydrogen/air. Therefore, the spontaneous wave velocity in methane/air is smaller for the same temperature gradients. Since the values of sound speeds and the Chapman-Jouguet velocities for hydrogen/air and methane/air are fairly close, one can expect that the minimum size of the hot spot, which can produce detonation will be at least ten times greater for methane/air compared to hydrogen/air.

Classification of possible modes of the propagating combustion wave inspired by the spontaneous wave initiated by a temperature gradient is similar to that described by Liberman et al. [37] for hydrogen/oxygen. The pressure waves generated during the exothermic stage of reaction can couple and evolve into a self-sustained detonation, or the coupling and synchronization and mutual amplification between the travelling shock wave and reaction front failed, resulting in the pressure waves running away ahead of the deflagration wave. The outcome depends on the gradient steepness and the relationship between the speed of the spontaneous wave at the point where its velocity reaches a minimum, $\min\{U_{sp}\}$ and the characteristic velocities of the problem: the laminar flame speed $U_f$, the speeds of sound at the points $T^*$ and $T_0$: $a_s(T^*)$ and $a_s(T_0)$, speeds at the Newman point $a_N$, at the Chapman-Jouguet point, $a_{CJ}$ and the velocity of Chapman-Jouguet detonation $U_{CJ}$. Because of the limited space, we consider only conditions under which the temperature gradient can initiate detonation.

### *6.1. Detonation initiation by temperature gradient in CH$_4$-air. One-step chemical model*

Figure 13 shows the necessary condition for the formation of detonation according Eq. (12) for a one-step model, for $P_0 = 1\text{bar}$, $T^* = 1800\text{K}$ and for different temperatures outside the gradient: $T_0 = 300$, 500, 700, and 1000K. $T^*_{cr} = 1500\,\text{K}$ corresponds to the location of the spontaneous wave, where earlier released in the reaction energy resulted in the first noticeable pressure peak.

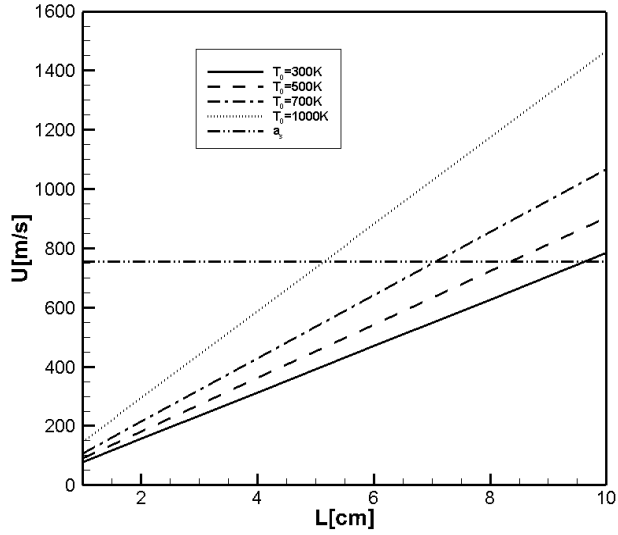

**Figure 13:** The intersection of lines $U_{sp}(T^*_{cr})$ with the sound speed corresponds to the steepest gradients producing detonation in $CH_4$/air for different $T_0$, computed for the one-step model.

The results of simulations for temperature gradients: $L = 7cm$ and $L = 9cm$, $P_0 = 1bar$, $T^* = 1800K$, $T_0 = 300K$, are shown in Figs. 14(a, b).

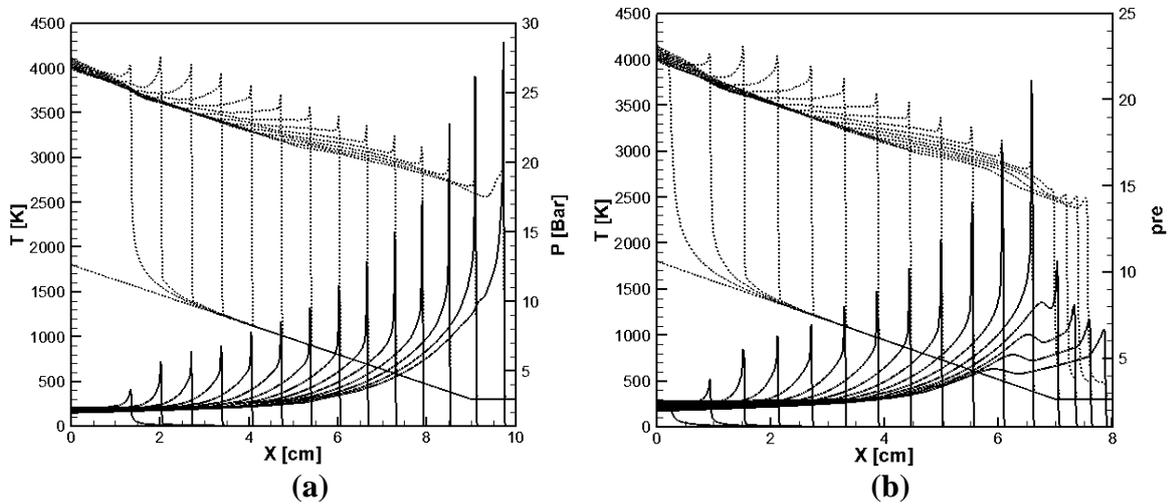

(a)        (b)

**Figure 14:** Evolution of the temperature (dashed lines) and pressure profiles (solid lines) during the formation of detonation at $P_0 = 1atm$. (a) $L = 9cm$; (b) $L = 7cm$.

According to the diagram in Fig.13 a steady detonation can be developed by the temperature gradient ($P_0 = 1bar$, $T^* = 1800K$, $T_0 = 300K$) if it steepness corresponds to $L \geq L_{cr} \approx 9cm$, which agrees well with Fig.14(a). In contrast to the scenario shown in Fig.

14(b), the developing detonation shown in Fig.14(a) $L = 7cm < L_{cr}$ quenches at $x \approx 6.7cm$. In this case the reactive zone starts to move slowly away from the leading shock wave. The rarefaction wave propagates into the reaction zone and the separation between the zone of heat release and the leading shock increases. As a result, the intensity of the shock wave becomes weaker and detonation quenches. For the steeper temperature gradient (Fig. 14b), the reaction velocity at the point, where the pressure wave overtakes the reaction wave, is not sufficient to sustain synchronous amplification of the pressure pulse in the flow behind the shock wave. As a result, the pressure wave runs ahead of the reaction wave and the velocity of the reaction wave decreases. The evolution of the reaction and pressure (shock) wave velocities for the conditions of Fig. 14(a) and Fig. 14(b) is shown in Figs. 15(a, b). The velocity of the reaction wave was calculated from the trajectory of the reaction front and the velocity of the pressure wave was calculated from the trajectory of the maximum pressure of the pressure wave profile.

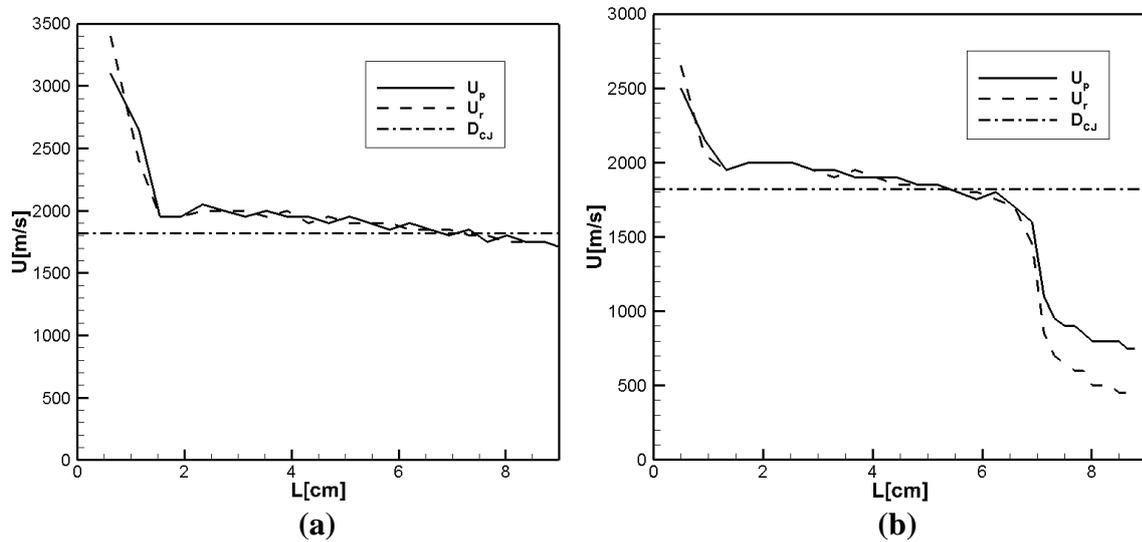

**Figure 15:** Velocities of the reaction wave (solid line) and pressure wave (dash-dotted lines) computed for the conditions in Fig. 14(a) and Fig. 14(b).

For the first time, phenomenon of spontaneous quenching of the developing detonation has been studied by He and Clavin [39, 40, 41] (see also Clavin and Searby [42]). This led to the definition of the critical size of the initial hot pocket of fresh mixture for ignition a detonation, observed by He & Clavin [39]. They also pointed that for a given temperature T* the critical temperature gradient for the spontaneous formation of a CJ detonation is defined by the local criterion $U_{sp}(x) \sim a*(x)$. It is seen that for the steeper gradient ($L = 7cm$), shown in Fig. 14(a), from the beginning the developed detonation follows closely the local CJ detonation in the early stage of propagation but later on it quickly quenches at $x > 6.5cm$.

He and Clavin [40, 41] also emphasized that the same temperature gradient, for which a detonation is quenching, can ignite a detonation for higher temperature $T_0$ outside the hot spot, as it is seen in Fig. 13. Indeed, it is shown in Fig. 16, that in agreement with Fig. 13, detonation does not quench and develops in a steady CJ detonation for the same size of the temperature gradient as in Fig. 14b, but for higher temperature $T_0 = 700K$. Radulescu et al. [46] noted that as more uniform reaction zone, as stronger reaction is coupled with the shock wave. According to Liberman et al. [37] at higher ambient temperatures the reaction front propagates at smoother ambient density, so that hydrodynamic resistance at the end and outside the gradient is smaller and the transition to detonation may occur for a steeper gradient (smaller $L_{cr}$).

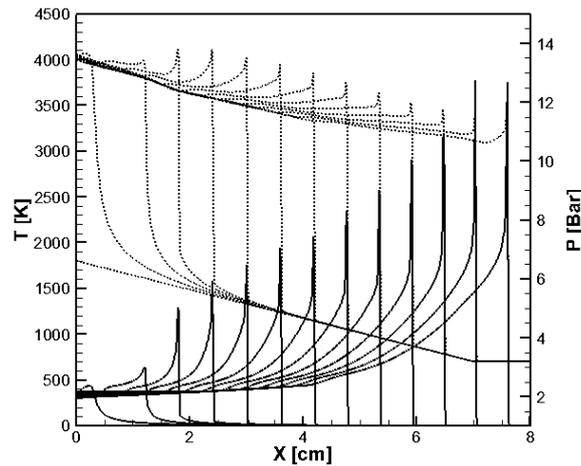

**Figure 16:** Evolution of the temperature (dashed lines) and pressure (solid lines) profiles during the detonation formation: $P_0 = 1atm$, $L = 7cm$, $T_0 = 700K$.

He and Clavin [40] explained spontaneous quenching of detonation in simple terms using a particular form of the quasi-steady-state approximation. In the classification of reaction waves initiated by temperature gradient by Liberman et al. [37] this corresponds to a quasistationary structure consisting of a shock wave and reaction zone, which may transform into a detonation propagating down the temperature gradient for the condition $a_N < \min\{U_{sp}\} < a_{CJ}$.

For a higher initial pressure the induction time decreases. Since $\left(\partial \tau_{ind} / \partial T\right)$ decreases rapidly with increasing pressure (approximating 10 times with increasing pressure from 1atm to 10atm), the spontaneous wave speed increases rapidly (Fig.12). Therefore, at high pressures the minimal steepness of the gradient necessary for the detonation initiation increases ($L_{cr}$ decreases). Evolution of the temperature and pressure profiles during the detonation development, calculated for the one-step model, at $P_0 = 5bar$ and $P_0 = 10bar$, $T_0 = 300K$ is shown in Fig. 17(a, b).

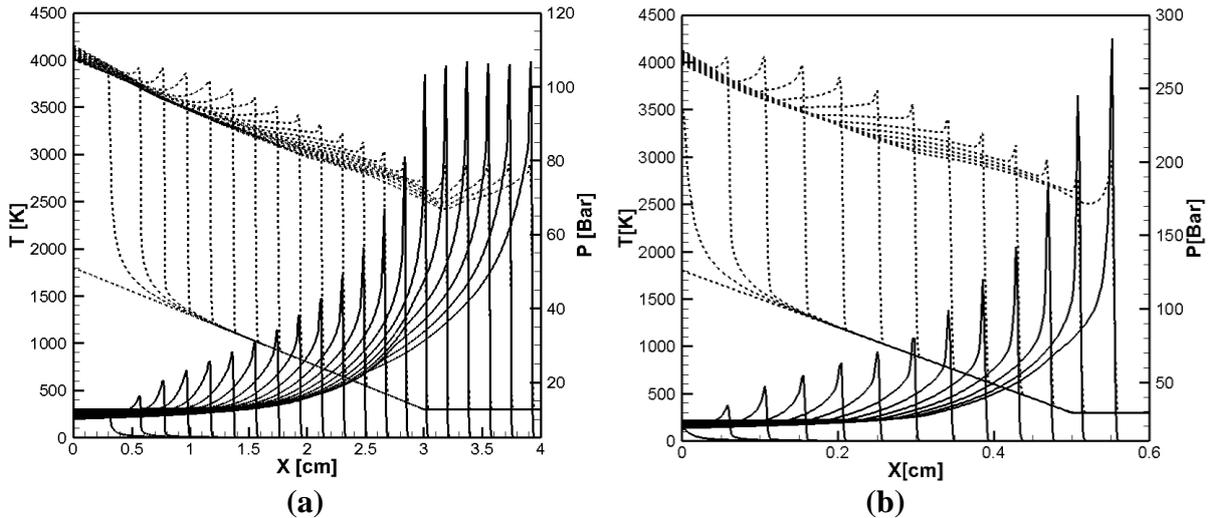

**Figure 17:** Evolution of the temperature (dashed lines) and pressure (solid lines) profiles during the formation of the detonation for the one-step model: (a) $P_0 = 5bar$; (b) $P_0 = 10bar$.

Since at high pressure the ranges of "speed" limits separating regions of different modes, which are determined by the sound speeds, $a_0$, $a*$ and $a_{CJ}$ decrease, the ranges for the realization of all combustion modes decrease correspondingly [37].

### 6.2. Detonation initiation by temperature gradient; detailed chemical model DRM-19

In this section we show that the scenario of detonation initiation by the temperature gradient in $CH_4$/air changes considerably for the detailed chemical model DRM19 compared to a one-step model. The ignition delay time of methane/air calculated using detailed chemical model and measured experimentally is much longer than that for the one-step model. At the same time $\left(\partial \tau_{ind} / \partial T\right)$ for the detailed chemical model DRM19 is about 2 orders larger than for the one-step model. This means that for "real chemistry" a successful detonation initiation through the temperature gradient requires much shallower gradient and much larger critical size $L_{cr}$ of the hot spot than that predicted by a one-step model.

In the case of detailed chemistry the initiating reactions proceed without heat release, and the gas-dynamic perturbations at the induction stage are very weak. The wave of exothermal reaction follows the spontaneous wave path with the delay determined by the time scale of termination reactions. Therefore, a steady detonation was not observed in simulations for shallower gradients up to $L < 30 cm$ for same initial conditions as for in Fig. 14(a) ($P_0 = 1 bar$, $T^* = 1800K$, $T_0 = 300K$), while the one-step model can yield successful detonation for much steeper gradients. Using the condition of Eq.(12) applied for the detailed DRM19 model, and the diagram similar to Fig.13, the minimum size of a hot spot, which can trigger a detonation for $T^* = 1800K$, $T_0 = 300K$ is estimated: 140cm at $P_0 = 1 bar$, 40cm at $P_0 = 5 bar$, 12cm at $P_0 = 10 bar$. An example, Fig.18 shows the formation of a combustion wave behind the weak shocks, which run ahead of the reaction wave front.

One would expect that the implementation of a steady detonation for the same temperature gradient can occur at higher pressures. However, the scenario for detailed chemistry at higher pressures differs significantly from that shown in Figs. 17(a, b) for a one-step chemistry. Figures 19(a, b) show the time evolution of temperature and pressure profiles for the same conditions as in Fig. 17(a, b). It is seen that the velocity of the spontaneous reaction wave at the minimum point, where the pressure wave overtakes the reaction wave, is not sufficient to sustain synchronous amplification of the pressure pulse in the flow behind the shock. The pressure wave runs ahead of the reaction wave, the velocity of the reaction wave decreases and the developing detonation quenches.

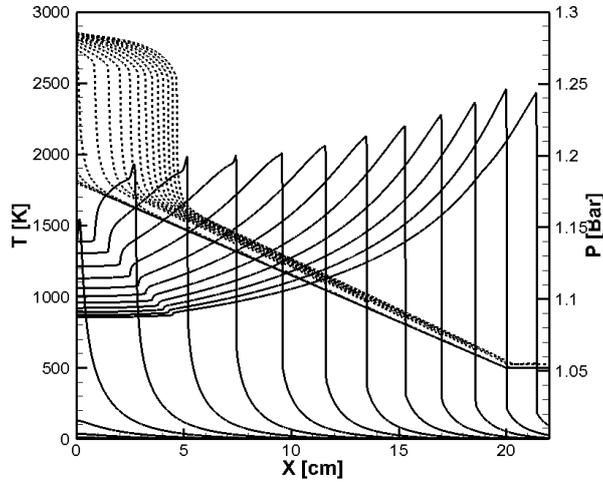

**Figure 18:** Evolution of the temperature (dashed lines) and pressure (solid lines) profiles calculated for DRM19 model for $T^*=1800K$, $T_0=300K$, $P_0=1bar$.

As it was discussed in the previous section, according He and Clavin [40, 41] Radulescu et al. [46] and Liberman et al. [37], the same temperature gradient, for which a detonation is quenching, can ignite a detonation for higher ambient temperature $T_0$. One should keep in mind, that Eq. (12) provides only the necessary but not sufficient condition for the minimum size $L_{cr}$ of the temperature gradient for producing a steady detonation.

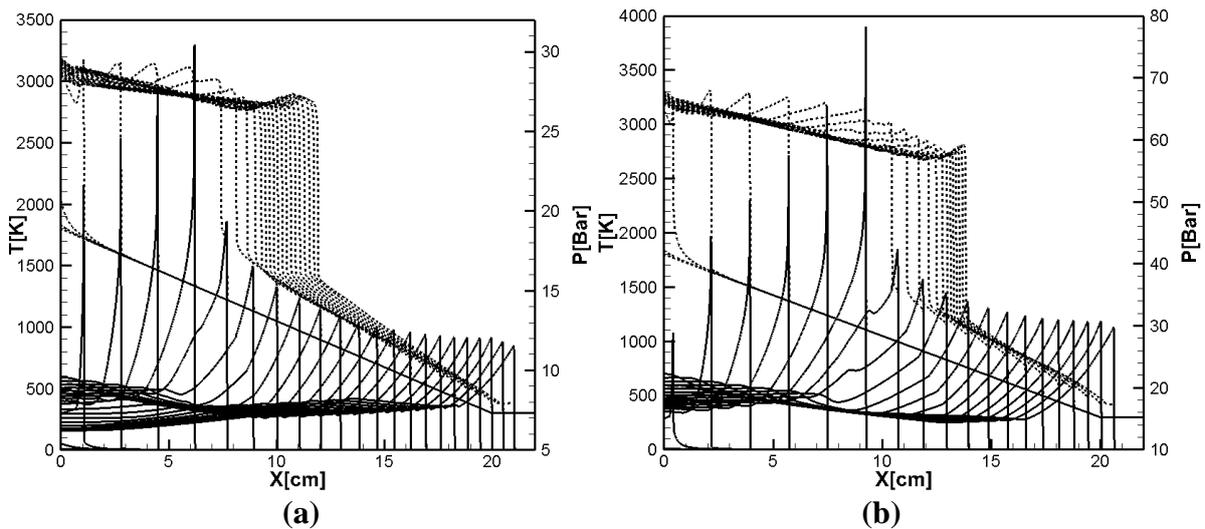

**Figure 19:** Evolution of the temperature (dashed lines) and pressure (solid lines) calculated for DRM-19: (a) $P_0=5bar$; (b) $P_0=10bar$; $T^*=1800K$, $T_0=300K$.

Although the expected sizes of the temperature gradient for initiation detonation predicted by Eq.(12) decrease with increasing initial pressure, it was found that the developing detonation quenches for all values $T_0 < 1100\,K$ in spite of rather large ($L > 20\,cm$) sizes of the hot spot.

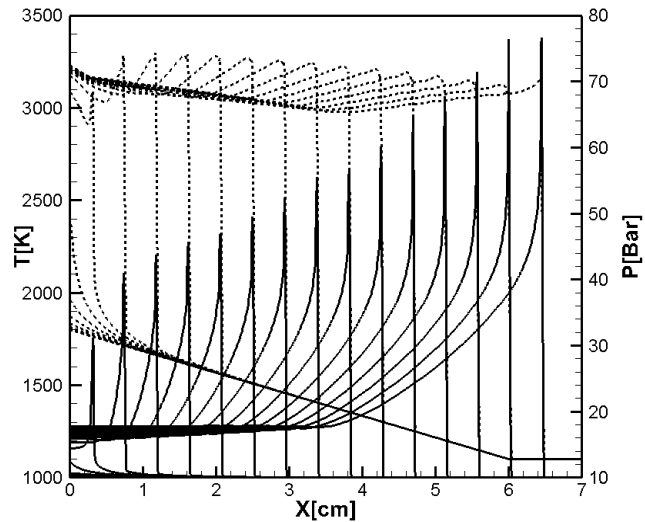

**Figure 20:** Evolution of the temperature (dashed lines) and pressure (solid lines) for developing a steady detonation calculated for DRM19 model; $P_0 = 10\,bar$, $T_0 = 1100\,K$.

Even at the initial pressure $P_0 = 10\,bar$ a steady detonation can be produced by the temperature gradient only at high ambient temperature $T_0 \geq 1100\,K$, as it is shown in Fig. 20, where a steady CJ-detonation is developed for a steeper gradient with the critical size $L_{cr} = 6\,cm$ of the hot spot. In a sense, the high ambient temperature outside the gradient region is equivalent, but not completely, to a shallower gradient. The induction stage, which is distinctive for real chemical reactions can be "skipped" at sufficiently high ambient temperatures, and the scenario of a detonation wave formation at the end of the gradient or outside the gradient becomes more complicated.

Thus, even at the relatively high pressure 10 atm, the minimum size of a hot spot necessary to initiate detonation by the gradient mechanism is much larger than for the same conditions for the highly reactive hydrogen/air. At very high pressures, 40-50atm, a steady detonation can be produced by a steeper gradient in a smaller hot spot, but also only at a sufficiently high temperature $T_0$ outside the hot spot, which is as higher as higher initial pressure. Another feature of the detonation initiation in methane/air is that since triggering detonation by the gradient mechanism is possible only at a high ambient temperatures, the volume thermal explosion ahead of the shock can occur before the developing detonation leaves the temperature gradient. Depending on the relation between the time of a steady detonation formation and the time of the thermal explosion development, the detonation can "meet" with a thermal explosion at the end or outside the temperature gradient. Therefore, such scenario in a certain sense looks more like a thermal explosion.

The difference between the detailed model and one-step model can be better understood using the thermal sensitivity of the induction time

$$\beta = -(T/\tau_{ind})(\partial \tau_{ind}/\partial T). \qquad (14)$$

Approximating the induction time by the exponent, $\tau_{ind} = A\exp(E_a/RT)$, and taking into account that $(\partial \tau_{ind}/\partial T) = -\tau_{ind}(E_a/RT^2)$, we obtain

$$\beta = (E_a/RT) \qquad (15)$$

where $E_a$ can be viewed as an effective global activation energy, which is different for the one-step and for the detailed chemical model.

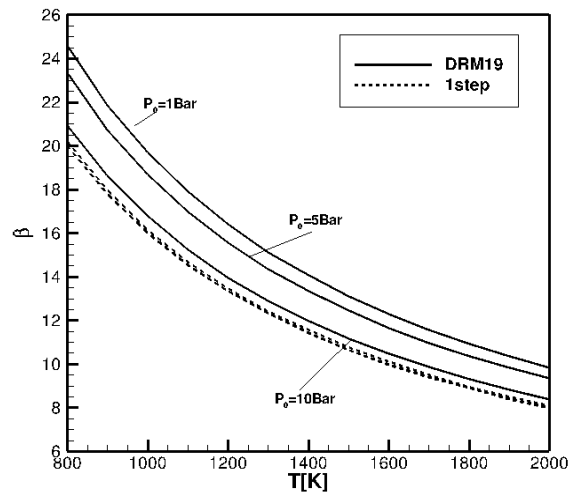

**Figure 21:** $\beta(T)$ calculated for the one-step model (dashed lines) and for the DRM19 model (solid lines) for different pressure: $P_0 = 1\text{bar}$, $P_0 = 5\text{bar}$, $P_0 = 10\text{bar}$.

Figure 21 shows β(T) calculated using the one-step model Eq. (1) for methane/air, and for the DRM19 model for different pressures: $P_0 = 1, 5,$ and $10\,\text{bar}$. From Fig.21 it is seen that $\beta(T)_{DRM}$ for the DRM19 model approaches $\beta(T)_{1-\text{step}}$ for the one-step model with increasing pressure. Thus, the global activation energy for single-stage and detailed models becomes almost the same at sufficiently high pressures. However, since for the detailed model both values of $\tau_{ind}$ and $(\partial \tau_{ind}/\partial T)$ remain at least two orders of magnitude smaller than they are for the one-step model, the minimum size of the temperature gradient is much larger in the case of real detailed chemistry than it is for the one-step model and the conditions for the detonation initiation are considerably complicated.

## 7. Discussions and Conclusions

In this work, we used detailed chemical kinetics models to study the conditions for a detonation initiation by the spontaneous wave inspired by a temperature gradient in highly reactive hydrogen/air mixtures and low reactive methane/air mixtures. Since many theoretical studies and simulations, undertaken to understand the explosion origin, suggest that the spontaneous wave and the gradient mechanism are the mechanism of the transition to detonation, specific focus has been placed on identifying conditions under which the detonation can be initiated by a spontaneous wave. The problem in question was studied using high-resolution numerical simulations, and the obtained results give the scales of the initial temperature nonuniformity and required for initiation a detonation by the temperature gradient.

The results of high resolution simulations performed for one-step models were compared with simulations for detailed chemical models. The calculated values of induction times for $H_2$/air and for $CH_4$/air were validated against experimental measurements for a wide range of temperatures and pressures. It is found that the requirements in terms of the temperature and size of hot spots, which produce a spontaneous wave which can initiate the detonation, are quantitatively and qualitatively different for one-step models compared to the detailed chemical models. The induction time and the temperature derivative of the induction time, which determines the speed of the spontaneous wave, are in orders of magnitude smaller for one-step models in comparison with the real values calculated using detailed models and measured in experiments. As a consequence, for the one-step models the hot spots are much smaller and the temperature gradients initiating a detonation much steeper than those calculated for detailed chemical models. The difference between the one-step and the detailed chemical model is more pronounced for low reactive $CH_4$/air mixture. In this case, even at a high pressure of 10 atm, the minimum size of the hot spot for which the spontaneous wave can initiate detonation exceeds 6 cm. Also, a temperature gradient can produce a spontaneous wave igniting detonation only for a high temperatures exceeding 1100K outside the hot spot. Such a temperature can trigger a thermal explosion and unlikely to be achieved during the flame acceleration prior to DDT. One of the conclusions is that the gradient mechanism of DDT, which was previously proposed on the basis of two-dimensional simulations using a one-step model, is unlikely to be a mechanism of DDT at least in the case of methane/air. Contrary to methane/air mixtures, at very high pressures the detonation in hydrogen/air can be ignited by a small-scale initial nonuniformity, which is of substantial practical interest for risk assessment to minimize accidental explosions, in particular, for safety guidelines in industry and nuclear power plants.

It is known [12] that for a flame propagating in a tube with no-slip walls viscous dissipation in the flow ahead of the advancing flame heats the gas mixture increasing temperature in the boundary layer by 200-300K higher than the gas temperature in the bulk flow near the tube axis. According to simulations of the flame propagating in submillimetre two-dimensional channels with a one-step model [55-58] viscous heating in the boundary layer induces local explosions near the boundary, which in turn leads to DDT. For a one-step model this leads to the autoignition producing a spontaneous wave, which occurs during the time about 10μs for submillimeter channels (about 60µs for 2cm channel) that elapse between the passage of the precursor shock and the flame arrival. However, the increased temperature and pressure in the boundary layer are far below the ignition threshold, and the ignition times are more than 100 times longer than the time that elapse between the passage of the shock and the arrival of the flame even at pressures up to 10 atm, which can be achieved prior to DDT for hydrogen/air and more than 1000 times longer for methane/air. Therefore, most likely, the viscous heating in the boundary layer cannot cause either autoignition or spontaneous wave to initiate DDT

Another shortcoming of one-step models is the incorrect velocity-pressure dependence predicted by one-step models. Figures 22 (a, b) show the velocity-pressure dependence of the laminar flames for $H_2$/air in Fig. 22(a) and for $CH_4$/air in Fig. 22(b) calculated for the one step models and for detailed chemical models. Symbols in Fig. 22 (a, b) show the experimental measurements of the laminar flame velocities for different pressures.

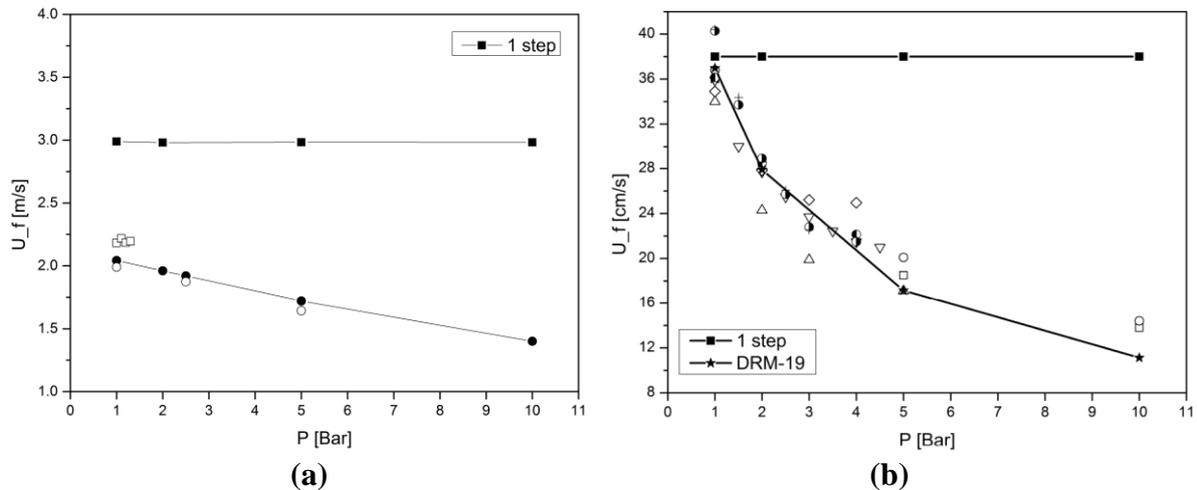

(a)          (b)

**Figure 22** The flame velocity-pressure dependence calculated for: one-step and detailed chemical models. (a): $H_2$/air; ● − detailed model [20]; experiments: □ - [21]; ○ - [47]. (b): $CH_4$/air; experiments: ▽ - [24], □ - [48], ○ - [49], △ - [50], ◇ - [51] ☻ - [52]; ☻ - [53]; +- [54].

The dependence of the flame velocity on pressure obtained for one-step models may cause incorrect values for the run-up distance predicted in the simulations of the flame acceleration and DDT.

The difference between induction times calculated using one-step models and detailed chemical models is inevitably associated with the calibration of one-step model parameters.

The adiabatic flame temperature, $T_b = Q/c_{p2} + T_1(c_{p1}/c_{p2})$ and the Chapman–Jouguet (CJ) velocity

$$D_{CJ} = \left(\frac{2\gamma_2^2 Q(\gamma_2 - 1)}{\gamma_2 + 1}\right)^{1/2} + \left\{\frac{2[(\gamma_2 - 1)Q + (\gamma_2 - \gamma_1)c_{v1}T_1]}{\gamma_2 + 1}\right\}^{1/2}. \qquad (16)$$

depends on the initial temperature, the specific heat ratios, $\gamma_{1,2}$, and the heat release $Q$. The one-step Arrhenius kinetics cannot exactly reproduce all properties of laminar flames and detonations. The parameters, $A$ and $E_a$ in the one-step model are calibrated for a particular fuel to give a reasonable approximation of the laminar flame speed at normal conditions. As a result, the global activation energy $E_a$ for one-step models is considerably smaller than the real activation energy for the detailed chemistry, which is demonstrated by comparison of the thermal sensitivity for $H_2$/air and $CH_4$/air in Fig.12 and Fig.21, correspondingly. For the same reasons the ratio of specific heats are taken 1.17 and 1.197 for $H_2$/air and $CH_4$/air in [17] and [8] instead of real values 1.39 $H_2$/air and 1.38 for $CH_4$/air.

All the same, simulations with a one-step model are of value as general examples of DDT in gaseous systems. However, it should be noted that, since both deflagration and detonation are stationary solutions, they appear to be stable attractors for all solutions in the vicinity of their base of initial data. This means that if the model allows unlimited acceleration of the flame, then the initially initiated deflagration will inevitably transit into detonation.

**Acknowledgments**

This work was sponsored in part by National Key R&D Program of China (No. 2017YFC0804700), by the National Natural Science Foundation of China under grants 11732003 and 11521062, and by funds of the opening project number KFJJ17-08M of State Key Laboratory of Explosion Science and Technology, BIT and in part (M.L.) by NORDITA and the Research Council of Norway under FRINATEK [grant No. 231444]. The authors are gratefully acknowledge insightful comments made by P. Clavin. M. Liberman acknowledges helpful discussions with many friends and colleagues, especially Grisha Sivashinsky and Igor Rogachevskii for their collaborations and deep insights into problems presented in this paper, and Alexander Konnov and Mikhail Kuznetsov for help with collection experimental data and for insights gleaned on detailed chemical schemes and DDT. Special thanks always to Dr. Koshatsky (Inna) for the wonderful environment, unfailing companionship and valuable advice.

**Disclosure statement**

**Appendix A: Verification of grid convergence: resolution and convergence tests**

In direct simulation of multi-species reactive flows, the flow and chemical time scales may be comparable, but the spatial resolution required for chemistry demands fine grids in order to resolve the thin reaction zones. The time step allowed when integrating the governing equations with explicit schemes is controlled by the diffusive stability limit dictated by these few chemical species. Resolution and convergence tests were thoroughly performed to ensure that the resolution is adequate to describe and to capture details of the problem in question and to avoid computational artifacts. This is especially important for a mixture consisting of many species with a large number of reactions in the case of a detailed chemical model. To verify the independence of the presented results on grid resolution, we performed large series of test simulations with uniform grid resolution for different grid sizes for normal and elevated pressures.

Figures A1 and A2 show the resolution and convergence tests for the structure and velocities of the laminar $H_2$/air flames for simulations with a one-step and detailed chemical schemes. The convergence of the solution is quite satisfactory already for 8 grid points per flame width at initial pressure 1atm.

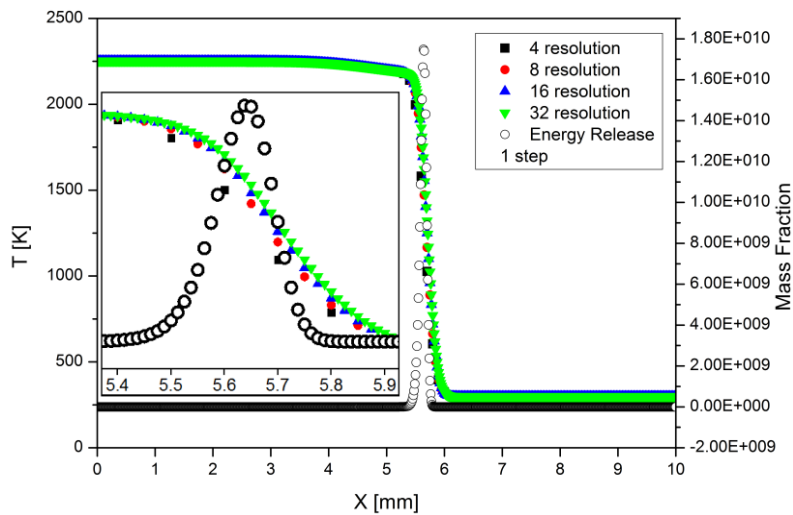

**Figure A1:** Resolution test for structure of stoichiometric hydrogen/oxygen flame at normal conditions ($T_0 = 300K$, $p_0 = 1atm$) for a one-step model.

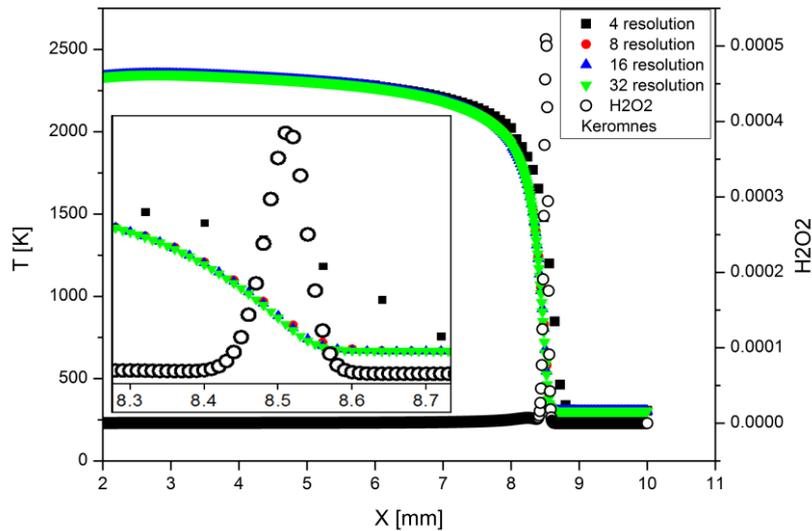

**Figure A2:** Resolution test for structure of stoichiometric hydrogen/oxygen flame at normal conditions ($T_0 = 300K$, $p_0 = 1atm$) for a detailed chemical model [20].

The convergence for the flame velocity in simulations with a one/step model and for detailed model [20] is shown in Fig. A3. A grid spacing independence was verified up to the resolution of 32 grid points per width of the flame and higher, which corresponds to the cell size less than 10-12μm. This resolution of 8 grid points per flame width is able to capture a good description of the flame at $P_0 = 1$atm, but higher resolutions are required at elevated pressures.

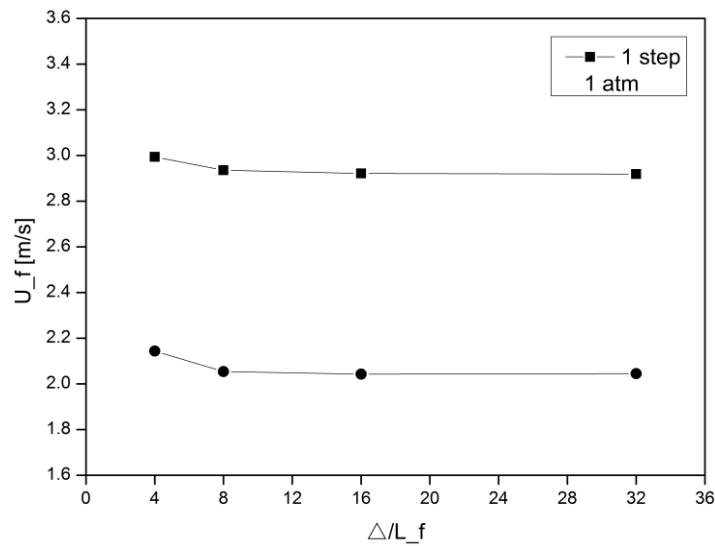

**Figure A3:** The convergence of solutions for the flame velocity in hydrogen/air flame at normal conditions for a one-step model and for a detailed chemical model [20].

Figure A4(a, b) shows resolution tests for the structure and velocity of methane-air flame for simulations with one-step and for DRM-19 chemical model.

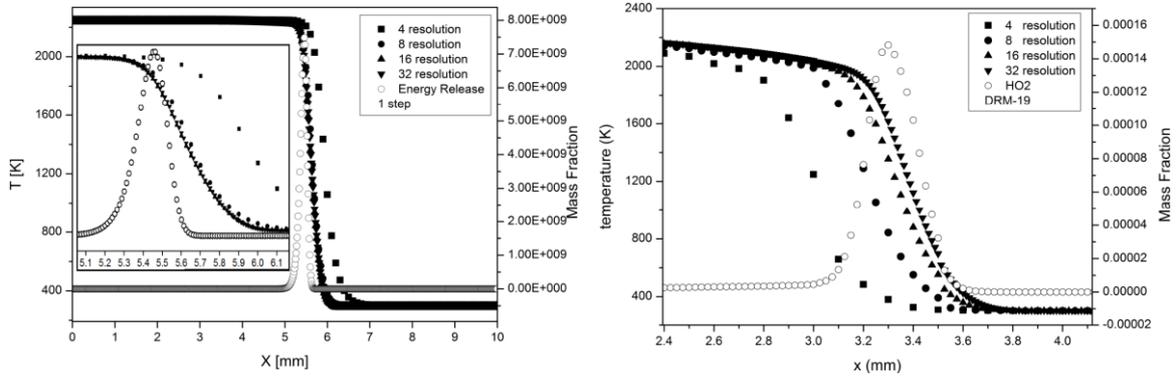

**Figure A4(a, b):** Resolution tests. a) One-step, b) DRM-19 models ($T_0 = 300$ K, $P_0 = 1$atm).

The most demanding is the case where the detonation arises as a result of auto-ignition inside the hot-spot. The detonation produced by the Zel'dovich gradient mechanism arose via the following stages: 1) the spontaneous combustion wave is formed, 2) the spontaneous wave decelerates and the pressure waves are formed behind the spontaneous wave, 3) the spontaneous reaction wave is coupled with pressure wave, which is amplified forming the shock wave, 4) detonation established after the transient process involving reaction wave accelerated in the flow behind the outgoing shock wave. Due to this sequence of the events one should appropriately resolve the combustion waves propagating through reacting medium on the background of elevated temperature and pressure behind the shock front. Besides the coupling of the reaction wave and shock should be resolved taking into account that the flame thickness is much larger than the width of the shock front. According to this we performed a common test for the accuracy of computational fluid codes, which was heavily investigated by Sod [59]. Figures A5(a, b) show solution for the Sod problem (Fig. 5a) and numerical solution to the problem (Fig. 5b).

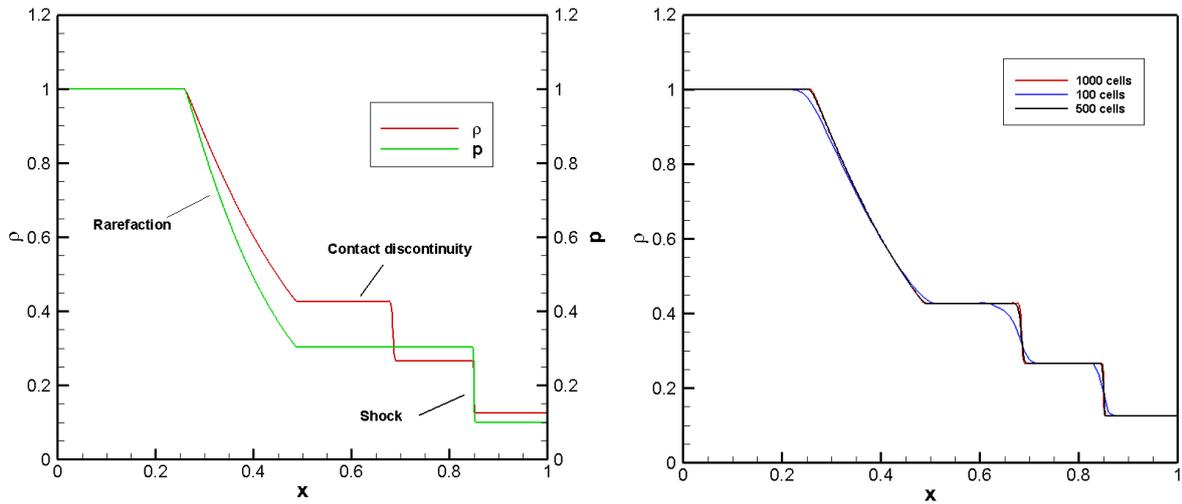

**Figure A5 (a, b):** Analytical solution (5a) for the Sod problem and numerical solution to the problem (Fig. 5b).

It is seen from Fig. A5b that the converged solution for the detonation initiation problem requires to use a finer resolution from the very beginning. For the model considered in the present paper the resolution was taken with at least 48 computational cells per flame front at normal conditions, that agree well with the results obtained previously. At elevated pressures the flame thickness decreases. On the other hand the diffusivity of the numerical scheme smoothens the shock front over 5 computational cells. Therefore no artificial coupling is possible for the chosen meshes determining fine resolution.

**List of figure captions**.

**Figure 1**: Induction times for stoichiometric hydrogen-air mixture at pressure: $P=1$ atm (a) and $P=2$ atm (b). one-step model (■), detailed model [20] (•). Experiments (a): □- Snyder et al. [26]; ○- Slack and Grillo [27]; ∆ – Hu et al. [28]; (b): □ - Slack and Grillo [27]; ○- Slack [25]; ∆- Bhaskaran and Gupta. [29].

**Figure 2**: Induction times for stoichiometric hydrogen-air mixture at $P=5$ atm calculated for the one-step and detailed chemical models. Experimental data are: □- Hu et al. [28]; ○- Wang et al. [30]. In Fig. 2(b) experimental data are: □- Hu et al. [30]; ○- Pan et al. [31].

**Figure 3**: Induction times for stoichiometric hydrogen-air mixture at pressures $P=10$ atm calculated for the one-step and detailed chemical models. Experimental data are: □- Hu et al. [28], ○- Pan et al. [31].

**Figure 4:** Induction times for methane-air at $P=1$ atm calculated for the one-step model, DRM19 and GRI3.0. Empty circles and squares are experimental data Zeng et al. [32] and Hu et al. [33], correspondingly.

**Figure 5**: Induction times for methane-air at $P=5$ atm. Empty circles and squares are experimental results from [32] and [33], correspondingly.

**Figure 6**: Induction times for methane-air at $P=10$ atm. Empty circles and squares are experimental results from [32] and [33], correspondingly.

**Figure 7:** Time evolution of the temperature (dashed lines) and pressure (solid lines) profiles during detonation initiation in $H_2$/air for one-step model. $P_0=1$ atm, $\Delta t = 2\mu s$.

**Figure 8:** Time evolution of the temperature (dashed lines) and pressure (solid lines) profiles during detonation initiation in $H_2$/air calculated for the detailed chemical model [20]; $P_0=1$ atm, $T^*=1500$ K, $\Delta t = 2\mu s$.

**Figure 9:** The minimum hot spot size $L=L_{cr}$ producing detonation at $P_0=1, 5, 10$ atm. (a): detailed model, $T_{cr}^*(1\text{atm})=1300$ K, $T_{cr}^*(5\text{atm})=1400$ K, $T_{cr}^*(10\text{atm})=1410$ K. (b): one-step model $T_{cr}^*(1\text{atm})=1200$ K, $T_{cr}^*(5\text{atm})=1300$ K, $T_{cr}^*(10\text{atm})=1400$ K.

**Figure 10:** Time evolution of the temperature (dashed lines) and pressure (solid lines) profiles during detonation initiation in $H_2$/air at $P_0=5$ atm, $T^*=1500$ K. (a): detailed model; (b): one-step model, $\Delta t = 2\mu s$.

**Figure 11:** Time evolution of the temperature and pressure profiles during detonation initiation in $H_2$/air at $P_0=10$ atm, $T^*=1500$ K. (a): detailed model; (b): one-step model.

**Figure 12**: $(\partial \tau_{ind}/\partial T)$ for $H_2$/air at initial pressure 1, 5 and 10 atm calculated for detailed (solid lines) and one-step (dashed lines) models.

**Figure 13:** The intersection of lines $U_{sp}(T_{cr}^*)$ with the sound speed corresponds to the steepest gradients with different $T_0$, producing detonation in $CH_4$/air for the one-step model.

**Figure 14:** Evolution of the temperature (dashed lines) and pressure profiles (solid lines) during the formation of detonation at $P_0 = 1atm$. (a) $L = 9cm$; (b) $L = 7cm$.

**Figure 15:** Velocities of the reaction wave (solid line) and pressure wave (dash-dotted lines) computed for the conditions in Fig. 14(a) and Fig. 14(b).

**Figure 16:** Evolution of the temperature (dashed lines) and pressure (solid lines) profiles during the detonation formation: $P_0 = 1atm$, $L = 7cm$, $T_0 = 700K$.

**Figure 17:** Evolution of the temperature (dashed lines) and pressure (solid lines) profiles during the formation of the detonation for the one-step model: (a) $P_0 = 5bar$; (b) $P_0 = 10bar$.

**Figure 18:** Evolution of the temperature (dashed lines) and pressure (solid lines) profiles calculated for DRM19 model for $T^* = 1800K$, $T_0 = 300K$, $P_0 = 1bar$.

**Figure 19:** Evolution of the temperature (dashed lines) and pressure (solid lines) calculated for DRM-19: (a) $P_0 = 5bar$; (b) $P_0 = 10bar$; $T^* = 1800K$, $T_0 = 300K$.

**Figure 20:** Evolution of the temperature (dashed lines) and pressure (solid lines) for developing a steady detonation calculated for DRM19 model; $P_0 = 10bar$, $T_0 = 1100K$.

**Figure 21:** β(T) calculated for the one-step model (dashed lines) and for DRM19 model (solid lines) for different pressure: $P_0 = 1bar$, $P_0 = 5bar$, $P_0 = 10bar$.

**Figure 22** The flame velocity-pressure dependence calculated for: one-step and detailed chemical models. (a): $H_2$/air; ● – detailed model [20]; experiments: □ - [21]; ○ - [47]. (b): $CH_4$/air; experiments: ▽ - [24], □ - [48], ○ - [49], △ - [50], ◇ - [51] ☾ - [52]; ◐ - [53]; +- [54].

**Figure A1:** Resolution test for structure of stoichiometric hydrogen/oxygen flame at normal conditions ($T_0 = 300K$, $p_0 = 1atm$) for a one-step model.

**Figure A2:** Resolution test for structure of stoichiometric hydrogen/oxygen flame at normal conditions ($T_0 = 300K$, $p_0 = 1atm$) for a detailed chemical model [20].

**Figure A3:** The convergence of solutions for the flame velocity in hydrogen/air flame at normal conditions for a one-step model and for a detailed chemical model [20].

**Figure A4(a, b):** Resolution tests. a) One-step, b) DRM-19 models ($T_0 = 300$ K, $P_0 = 1$atm).

**Figure A5 (a, b):** Analytical solution (5a) for the Sod problem and numerical solution to the problem (Fig. 5b).